\documentclass[a4paper,fleqn]{cas-sc}

\usepackage[numbers]{natbib}
\usepackage{setspace}
\usepackage{lineno}
\usepackage{comment}

\def\tsc#1{\csdef{#1}{\textsc{\lowercase{#1}}\xspace}}
\tsc{WGM}
\tsc{QE}
\tsc{EP}
\tsc{PMS}
\tsc{BEC}
\tsc{DE}
\begin{document}
\let\WriteBookmarks\relax
\def\floatpagepagefraction{1}
\def\textpagefraction{.001}
\shorttitle{Stochastic resonance in adaptive dynamical networks}
\shortauthors{V.Semenov}

%\title [mode = title]{L{\'e}vy noise-induced wavefront propagation for bistable media, ensembles and delayed-feedback oscillators}                      

\title [mode = title]{Stochastic resonance in adaptive dynamical networks}                      

\author[1]{Vladimir V. Semenov}[orcid=0000-0002-4534-8065]
\corref{cor1}
\ead[url]{semenov.v.v.ssu@gmail.com}
\address[1]{Institute of Physics, Saratov State University, 83 Astrakhanskaya str., 410012 Saratov, Russia}

\cortext[cor1]{Corresponding author}

\begin{abstract}
The role of adaptive coupling in controlling collective noise-induced dynamics is demonstrated on an example of stochastic resonance in an ensemble of coupled overdamped bistable oscillators. Specifically, it is found that tuning the adaptive coupling parameters allows one to either enhance or suppress stochastic resonance and to shift the optimal noise intensity that yields the most regular stochastic oscillations. All the revealed effects are studied in numerical simulations for two kinds of adaptive coupling models. The first one is a simplified, phenomenological model designed such that the coupling becomes stronger for oscillators with larger differences between their dynamical variables (two configurations for local interaction and one option for global coupling are under consideration). The second one is a local memristive coupling model that provides a physically motivated realization of adaptive interactions with similar properties. In all the considered cases, adaptive coupling is shown to be a flexible stochastic resonance control tool, both in the absence and in the presence of the static coupling component.
\end{abstract}

%\begin{highlights}
%\end{highlights}


\maketitle

\doublespacing

\section{Introduction}
\label{intro}
Stochastic resonance \cite{gammaitoni1998,anishchenko1999} is observed across an extremely wide variety of stochastic systems, including models for the periodicity of ice ages \cite{nicolis1981,nicolis1982,benzi1982} and the anomalous warming of equatorial waters of the Pacific \cite{stone1998}, lasers \cite{mcnamara1988,fioretti1993}, electronic circuits \cite{luchinsky1999-1,luchinsky1999-2,calvo2006}, chemical reactions \cite{leonard1994,foerster1996,guderian1996,hohmann1996,yang1999}, the mammalian auditory system \cite{hong2006} and many other examples. A list of stochastic resonance systems is not limited to single oscillators, but also encompasses bistable media \cite{benzi1985,jung1995,wio1996,wio2002,lai2009}, dynamical networks \cite{gosak2011,tang2012,semenov2022,semenov2026} and time-delay oscillators considered as spatially extended systems evolving in quasi-space \cite{semenov2016,semenov2025}. Furthermore, the occurrence of stochastic resonance can be accompanied by additional effects, for instance, spiral wave excitation \cite{jung1995} and chimera states \cite{semenov2016}.

The scientific importance of controlling stochastic resonance stems from the broad range of dynamical systems exhibiting this phenomenon and its wide range of practical applications (for instance, see the ones listed in review \cite{gammaitoni1998}). It includes improved energy harvesting \cite{zheng2014} and information transmission \cite{longtin1991} approaches, enhanced detector sensitivity, ranging from the Josephson escape detector \cite{ladeynov2023} to acoustic sensors \cite{shuyao2016}, upgraded pattern fabrication for semiconductor and optoelectronic industries \cite{roy2019} and machinery fault detection \cite{qiao2019}. Depending on the requirements of a particular practical application, it may be necessary to enhance or suppress stochastic resonance or to shift the optimal noise intensity to a more suitable range. For these purposes, different strategies can be employed: introducing time-delayed feedback \cite{mei2009,jia2010}, varying the noise's correlation time \cite{haenggi1993} and the probability density function \cite{dybiec2006,dybiec2009}, simultaneous applying additive and multiplicative noise \cite{qiao2016}. In systems of coupled oscillators, control of stochastic resonance can be achieved by varying the coupling strength or by modifying the coupling topology \cite{neiman1995,nicolis2017,semenov2022,semenov2026}. The coupling-based approach is considered in the present paper, with a particular focus on the impact of adaptive coupling properties.

Adaptive coupling is known to essentially change the network dynamics as compared to the effects in the presence of static interactions \cite{gross2008,gross2009,berner2023}. Among the reported effects are enhanced synchronization \cite{zhou2006,ren2007,zhu2010} and control of the synchronized oscillation characteristics \cite{ren2007}, explosive synchronization \cite{zhang2015}, formation of frequency clusters  \cite{roehr2019,berner2019,berner2019-2}, chimera \cite{kasatkin2018,kasatkin2018-2,kasatkin2019} and solitary \cite{berner2020} states. In the context of stochastic phenomena, adaptive coupling is found to interplay with noise such that different types of noise-induced switching are realized in the same coupled excitable oscillator model \cite{bacic2018}. Furthermore, when the coupling strength is adaptive, stochastic resonance in small-world networks of bistable elements is enhanced \cite{wu2011}. 

Considering different coupling nonlinearities and implementing adaptive interactions in various forms, the present study complements the results of paper \cite{wu2011} and demonstrates that the coupling adaptivity can lead to a diverse range of effects beyond the stochastic resonance enhancement. Moreover, one of the studied coupling models involves the interaction through the memristor. Memristive coupling is capable of supporting travelling waves and controlling their characteristics \cite{korneev2024}. Employing a memristor as a coupling element enables the observation of initial-condition-dependent synchronization in regular \cite{korneev2020} and chaotic \cite{korneev2021-3} self-oscillators, as well as in wave processes occurring in single- \cite{korneev2021} and multilayer \cite{korneev2022} networks. Here, the range of phenomena associated with the action of memristive coupling is extended by demonstrating its impact on stochastic resonance.

\section{Model and methods}
\label{model_and_methods}
The systems considered in this study are networks of overdamped bistable oscillators characterized by the coexistence of two stable steady states. Each oscillator has its own statistically independent source of additive noise and is subject to a common external periodic force. The governing equations are studied in the following form:
\begin{equation}
\label{eq:general}
\begin{array}{l}
\dfrac{dx_{i}}{dt}=mx_{i}-x_{i}^3+A\sin(\omega_{\text{e}} t)
+\sqrt{2D}n_i(t)+f_i(\vec{x},\vec{\sigma}). 
\end{array}
\end{equation}
Here, $x_i$ denotes the dynamical variable of the $i$-th oscillator, where $i=1,2,\ldots,N$ and $N$ is the total number of oscillators, while $f(\vec{x},\vec{\sigma})$ represents the coupling term depending on both oscillators' dynamical variables, $\vec{x}$, and the instantaneous coupling element states $\vec{\sigma}$. Throughout this study, all the networks comprise $N=100$ elements. Parameter $m$ governs the dynamics of an individual network element, determining whether it is monostable ($m<0$) or bistable ($m>0$). In the present study, all the network elements are bistable with $m=0.25$, corresponding to the coexistence of two stable steady states, $x^*_{1,2}=\pm0.5$, in the phase space of a single oscillator. Terms $\sqrt{2D}n_i(t)\in\mathbb{R}$ represent Gaussian white noise of intensity $D$, i.e., $<n_i(t)>=0$ and $<n_i(t)n_{j}(t)>=\delta_{ij}\delta(t-t')$, $\forall i,j$. The external forcing amplitude is fixed at $A=0.04$, below the threshold required to induce transitions between the states $x^*_{1,2}$ in the ensembles without noise.\footnote{It is worth noting that stochastic resonance can also be observed when the driving amplitude significantly exceeds the threshold value \cite{pankratov2002}.} The forcing frequency is chosen to be sufficiently low, $\omega_{\text{e}}=0.005$, to observe a well-pronounced stochastic resonance effect.

Networks (\ref{eq:general}) are investigated numerically by integrating the equations using the Heun method \cite{mannella2002}, with a time step of $\Delta t=0.001$ and a total integration time of $t_{\text{total}}=10^6$. The initial conditions are random and uniformly distributed over the interval $x_i(t=0)\in[-0.75,0.75]$.
The simulation results are visualized using space-time plots $x_i(t)$ and time realizations $\overline{x}(t)$  and $\overline{\sigma}(t)$ which are averaged over the ensemble values of dynamical variables $x_i$ and $\sigma_i$, $\overline{x}(t)=\dfrac{1}{N}\sum\limits_{i=1}^{N}x_i(t)$ and $\overline{\sigma}(t)=\dfrac{1}{N}\sum\limits_{i=1}^{N}\sigma_i(t)$. 
Quantities $\overline{x}$ and $\overline{\sigma}$ are introduced to characterize the instantaneous global state of the ensemble. In addition, the time-averaged value of $<\overline{\sigma}(t)>$ is taken into account. In fact, $<\overline{\sigma}(t)>$ are values of $\sigma_i(t)$ averaged over both time and ensemble.

Following Refs. \cite{semenov2022,semenov2026}, the ensemble-averaged power spectrum is used to globally characterize stochastic resonance: $\overline{S}(\omega)=\dfrac{1}{N}\sum\limits_{i=1}^N S_i(\omega)$, where $S_i(\omega)$ denotes the power spectrum of the oscillations $x_i(t)$. This approach allows one to highlight the similarity between the effects observed in ensembles and stochastic resonance in a single oscillator. As the noise intensity increases, the evolution of the ensemble-averaged power spectrum reproduces the characteristic features of classical stochastic resonance (see Figs. 1 in Refs. \cite{semenov2022,semenov2026} and Fig.~\ref{fig2} in the present paper).

In addition, the signal-to-noise ratio (SNR) is used as a quantitative measure of the noise-induced dynamics regularity and its variation with the coupling parameters and noise intensity. Here, the SNR is defined following the approach commonly used in radiophysics and electronics. The power spectrum of each network oscillator exhibits a spectral peak $S_{\text{max}}$ at the frequency of the external forcing, which corresponds to the dominant spectral component. In the vicinity of this peak, the spectrum also has a minimum $S_{\text{min}}$. The quantities $S_{\text{max}}$ and $S_{\text{min}}$ are schematically illustrated in Fig.~\ref{fig2}~(e) using the ensemble-averaged spectrum as an example. More specifically, $S_{\text{min}}$ is calculated as the mean value of the spectral components $S(\omega)$ over the frequency intervals [$0.75\omega_{\text{e}}:0.85\omega_{\text{e}}$] and [$1.15\omega_{\text{e}}:1.25\omega_{\text{e}}$], which exclude the vicinity of the main spectral peak at the external forcing frequency $\omega_{\text{e}}$. A commonly used definition of the SNR is $\text{SNR}=P_{\text{S}}/P_{\text{N}}$, where $P_{\text{S}}$ and $P_{\text{N}}$ denote the signal and noise powers, respectively. For a harmonic signal measured experimentally, the SNR can be expressed as $\text{SNR}=H_{\text{S}}/H_{\text{N}}$, where $H_{\text{S}}$ is the height of the spectral line above the background noise level and $H_{\text{N}}$ is the background noise level in the vicinity of the resonance frequency $\omega_{\text{e}}$. Accordingly, in terms of the power spectrum, the SNR is given by $\text{SNR}=S_{\text{max}}-S_{\text{min}}/S_{\text{min}}$. Considering the SNR as a function of the noise intensity $D$ yields a non-monotonic dependence that is characteristic of stochastic resonance, with an optimal noise intensity corresponding to a maximum of the SNR. To characterize the collective dynamics of the ensembles described by Eqs.~(\ref{eq:general}), the power spectrum of each oscillator is calculated from the corresponding time realization $x_i(t)$, followed by the evaluation of the SNR. The resulting SNR values are then averaged to obtain the ensemble-averaged signal-to-noise ratio, $\overline{\text{SNR}}$.

\section{Phenomenological coupling models}
\label{sec_phenomenomenological}
\subsection{Local interaction}
\label{sec_phenom_local}
The study begins from rings of locally coupled oscillators. Following Refs. \cite{zhou2006,zhu2010} where adaptive coupling is shown to enhance synchronization, the coupling terms are introduced such that the coupling strength increases proportionally to the difference between instantaneous values of the neighbour oscillator dynamical variables in order to suppress it. Two options for the coupling are under consideration (schematically illustrated in Fig.~\ref{fig1}). In the first case (see Eqs. (\ref{eq:local_individual}) and Fig.~\ref{fig1}~(a)), the coupling between $i$-th and $i+1$-th network nodes is realized through  $i$-th coupling element identically acting $i$-th and $i+1$-th oscillators. Each oscillator is coupled to its left and right neighbours separately:
%%%%%%%%%%%%%%%%%%%%%%%% FIG 1 %%%%%%%%%%%
\begin{figure}
\centering
\includegraphics[width=0.53\textwidth]{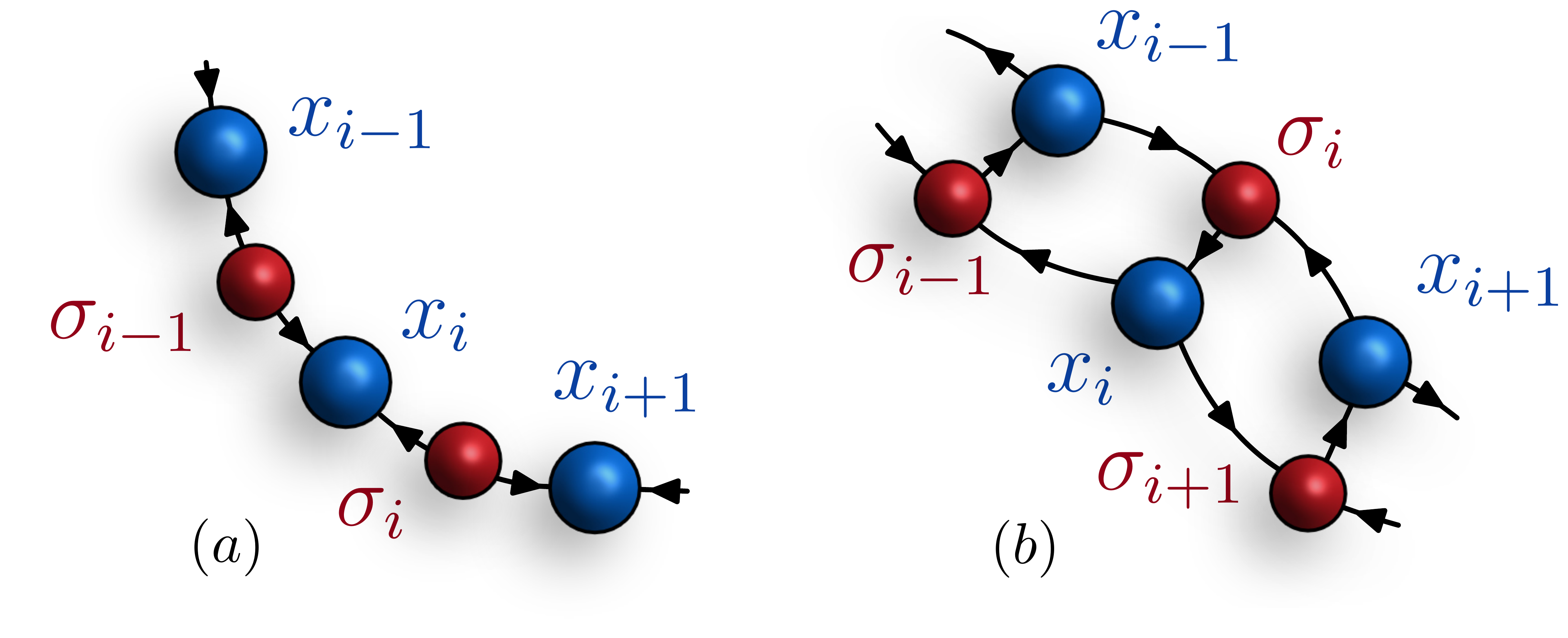}
\caption{Schematic illustration of the local coupling governed by Eqs. (\ref{eq:local_individual}) (panel (a)) and Eqs. (\ref{eq:local_summary}) (panel (b)).}
\label{fig1}
\end{figure}
%%%%%%%%%%%%%%%%%%%%%%%%%%%%%%%%%%%%%%
\begin{equation}
\label{eq:local_individual}
\begin{array}{l}
f_i(\vec{x},\vec{\sigma})= \dfrac{1}{2} \Bigl\{ (\sigma_{i-1}+\sigma_s)(x_{i-1}-x_i) 
+(\sigma_{i}+\sigma_{\text{s}})(x_{i+1}-x_i) \Bigr\}, \quad
\dfrac{d\sigma_{i}}{dt}=\dfrac{\gamma \Delta_i}{1+\Delta_i} - \alpha \sigma_i, \quad
\Delta_i=\left| x_i-x_{i+1} \right|.
\end{array}
\end{equation}
In the second coupling model (see Eqs. (\ref{eq:local_summary}) and Fig.~\ref{fig1}~(b)), the strength to a node $i$ from its both neighbors is the same and is defined by the summary dynamical variable difference $x_i - \dfrac{1}{2} \left\{ x_{i-1} + x_{i+1} \right\}$. In such a case, each oscillator interacts through one coupling element:
\begin{equation}
\label{eq:local_summary}
\begin{array}{l}
f_i(\vec{x},\vec{\sigma})= \dfrac{(\sigma_{i}+\sigma_{\text{s}})}{2} \Bigl\{ x_{i+1}+x_{i-1}-2x_i \Bigr\}, \quad
\dfrac{d\sigma_{i}}{dt}=\dfrac{ \gamma \Delta_i }{ 1+\Delta_i } - \alpha \sigma_i, \quad
\Delta_i=\left| x_i - \dfrac{1}{2} \left\{ x_{i-1} + x_{i+1} \right\} \right|.
\end{array}
\end{equation}
In both cases, coupling strength contains the static component determined by parameter $\sigma_s\geq 0$ being the same for all the oscillators and the adaptive one, $\sigma_i$, governed by the corresponding differential equation. In contrast to Refs. \cite{zhou2006,zhu2010,wu2011} the coupling element state equations of models (\ref{eq:local_individual}) and (\ref{eq:local_summary}) contain the decay term $-\alpha \sigma_i$. It protects against extreme multistability and continuous dependence of the oscillation characteristics on the initial conditions (are discussed below in the context of memristive coupling) and provides for observation of stationary stochastic processes. In the absence of the decay term, the adaptive coupling models imply the coupling strength growth over time (the right-hand side function of the coupling state equations takes values from zero to unity) or stabilization in the limit case $\Delta_i=0$ (cannot be achieved in the presence of individual noise sources in each oscillator). Such unbounded growth of the coupling strength is questionable from the perspective of real-world dynamical systems and does not seem to be an appropriate model for description of the real processes. In the presence of decay term, a balance between increasing the coupling strength (when the noticeable difference between the dynamical variable values takes place) and decreasing the interaction intensity (when the 'in-phase' oscillations are achieved) is organized.

%%%%%%%%%%%%%%%%%%%%%%%% FIG 2 %%%%%%%%%%%
\begin{figure}
\centering
\includegraphics[width=1.00\textwidth]{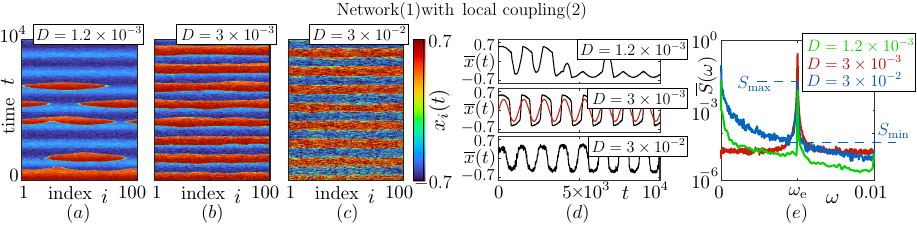}
\caption{Evolution of the stochastic resonance dynamics in network (\ref{eq:general}) with local coupling governed by Eqs. (\ref{eq:local_individual}) as the noise intensity $D$ increases: space-time plots (panels (a)-(c)), time realizations of the ensemble-averaged dynamical variable $\overline{x}(t)$ (panel (d)), and transformation of the  ensemble-averaged power spectrum $\overline{S}(\omega)$ (panel (e)). The red curve in the central inset in panel (d) illustrates the external forcing signal. Parameters are $m=0.25$, $N=100$, $A=0.04$, $\omega_{\text{e}}=0.005$, $\alpha=\gamma=10^{-3}$, $\sigma_{\text{s}}=0$.}
\label{fig2}
\end{figure}
%%%%%%%%%%%%%%%%%%%%%%%%%%%%%%%%%%%%%%

Consider network (\ref{eq:general}) in the presence of local coupling (\ref{eq:local_individual}) at zero static component of the coupling strength, $\sigma_s=0$, and positive fixed parameters $\gamma=\alpha=10^{-3}$. For sufficiently strong coupling and weak noise, system (\ref{eq:general}) exhibits coherent oscillations with rare or, moreover, absent transitions between two stable steady states $x_{1,2}^{*}$ [Fig.~\ref{fig2}~(a)]. As the noise intensity increases, transitions between the steady states become more and more frequent, while spatial coherence persists, as illustrated in the space-time plot in Fig.~\ref{fig2}~(b),(c). At an appropriate noise intensity $D_{\text{opt}}$, the transitions become nearly regular [Fig.~\ref{fig2}~(b)], with the optimal noise level depending on the coupling strength. Further increase in the noise intensity reduces the regularity of the system response [Fig.~\ref{fig2}~(c)]. These changes are also reflected in the time realizations of the ensemble-averaged variable $\overline{x}(t)$ as the noise intensity $D$ increases. In particular, oscillatory process $\overline{x}(t)$ exhibits the characteristic features of stochastic resonance in a single bistable system: transitions between the basins of attraction of the steady states $x_{1,2}^{*}$ become increasingly frequent with increasing noise intensity, while at $D=D_{\text{opt}}$ the oscillations become synchronized with the external periodic forcing. This behavior is illustrated in Fig.~\ref{fig2}~(d), where the external forcing is shown in the middle panel corresponding to $D=D_{\text{opt}}=3\times10^{-3}$. 

The similarity between stochastic resonance in the network and that observed in a single bistable oscillator is further demonstrated by the evolution of the ensemble-averaged power spectrum $\overline{S}(\omega)$, which exhibits the characteristic features of classical stochastic resonance [Fig.~\ref{fig2}~(e)]. As the noise intensity increases, the spectral peak at the frequency of the external forcing gradually increases in height and becomes most pronounced at an optimal noise intensity $D_{\text{opt}}$. With a further increase in the noise intensity, the response of the system to the external periodic forcing becomes less regular, resulting in the inverse evolution of the power spectrum.

\begin{figure}
\centering
\includegraphics[width=1.00\textwidth]{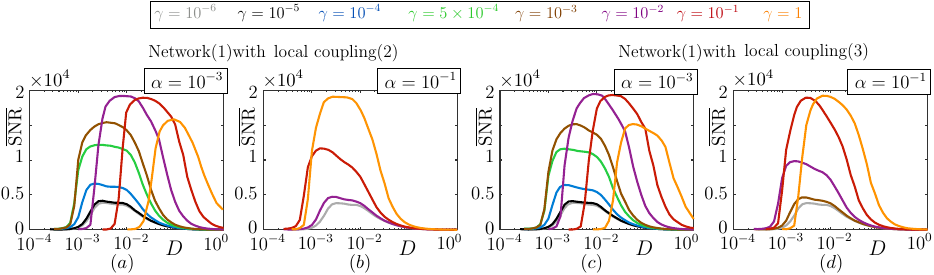}
\caption{Evolution of the stochastic resonance curves $\overline{\text{SNR}}(D)$ when varying adaptivity parameter $\gamma$ for fixed coupling decay parameter $\alpha=0.001$ (panels (a) and (c)) and $\alpha=0.1$ (panel (b) and (d)) in network (\ref{eq:general}) with local adaptive coupling governed by Eqs. (\ref{eq:local_individual}) (panels (a) and (b)) and Eqs. (\ref{eq:local_summary}) (panels (c) and (d)). The rest of the parameters are the same as in the previous figure.}
\label{fig3}
\end{figure}

The ensemble-averaged signal-to-noise ratio $\overline{\text{SNR}}$ is calculated from the power spectra of the individual oscillators. The resulting $\overline{\text{SNR}}(D)$ curves obtained for different values of parameter $\gamma$ reproduce the characteristic non-monotonic dependence associated with classical stochastic resonance in a single bistable oscillator. As shown in Fig.~\ref{fig3}~(a), increasing the parameter $\gamma$ can enhance stochastic resonance, as indicated by an increase in the maximum value of the SNR. However, this coupling-induced enhancement first achieves its maximum at $\gamma \approx 10^{-2}$ and then changes to the regularity suppression.  This effect is accompanied by significant horizontal shift of the SNR$(D)$ curves toward higher noise intensities. Thus, varying the coupling adaptivity parameter $\gamma$ provides an effective means of shifting the optimal noise intensity $D_{\text{opt}}$ over a broad range.  In particular, $D_{\text{opt}}=2\times 10^{-3}$ for $\gamma=10^{-4}$, whereas $D_{\text{opt}}=8\times 10^{-2}$ for $\gamma=1$ (see Fig.~\ref{fig3}~(a)).

Increasing the coupling decay parameter $\alpha$ does not change the stochastic resonance control qualitatively [Fig.\ref{fig3}~(a),(b)]. However, a quantitative difference is observed: the larger the value of $\alpha$, the higher the value of $\gamma$ required to achieve a comparable enhancement of stochastic resonance. The same behavior is observed when the coupling model is changed from Eqs.~(\ref{eq:local_individual}) to Eqs.~(\ref{eq:local_summary}), as demonstrated by the evolution of the SNR$(D)$ curves in Fig.~\ref{fig3}~(a),(b) and Fig.~\ref{fig3}~(c),(d). 
%Moreover, the qualitative features of stochastic resonance control remain unchanged when the coupling-element state equations is modified to $\dfrac{d\sigma_{i}}{dt}= \gamma \Delta_i^2 - \alpha \sigma_i$. This model likewise implies an increase in the coupling strength with increasing spatial incoherence, but, in contrast to Eqs.~(\ref{eq:local_individual}) and (\ref{eq:local_summary}), the right-hand-side function is not bounded within the interval $[0,1]$. The corresponding results are not shown, as they essentially reproduce those presented in Fig.~\ref{fig3}. 
Thus, the observations indicate that the properties of stochastic resonance in the presence of local adaptive coupling, whose strength increases with the difference between the dynamical variables, do not depend significantly on the way the adaptive coupling is introduced.
%, as discussed here on examples of coupling models Eqs.~(\ref{eq:local_individual}) and (\ref{eq:local_summary}).

%%%%%%%%%%%%%%%%%%%%%%%% FIG 3 %%%%%%%%%%%
%%%%%%%%%%%%%%%%%%%%%%%%%%%%%%%%%%%%%%

\begin{figure}
\centering
\includegraphics[width=0.95\textwidth]{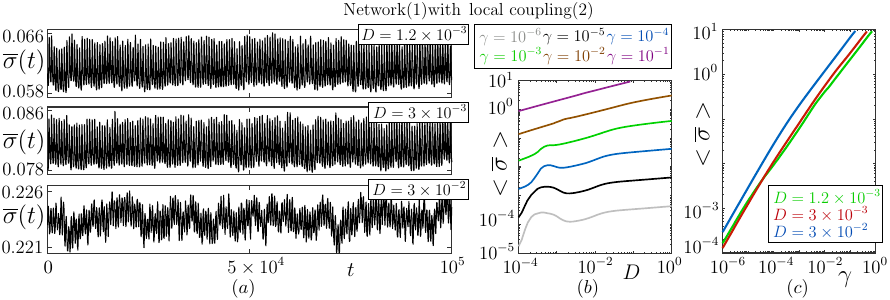}
\caption{Stochastic resonance dynamics in adaptive dynamical network (\ref{eq:general}) with local coupling governed by Eqs. (\ref{eq:local_individual}): evolution of time realizations $\overline{\sigma}(t)$ as the noise intensity $D$ increases; (b)-(c) Dependence of the mean value of time realizations $\overline{\sigma}(t)$ when changing noise intensity $D$ at various fixed $\gamma$ (panel (b)) and when varying $\gamma$ at different fixed noise intensity $D$ (panel (c)). The coupling decay parameter is $\alpha=0.001$. The rest of the parameters are the same as in previous figures.}
\label{fig4}
\end{figure}

\begin{figure}
\centering
\includegraphics[width=0.63\textwidth]{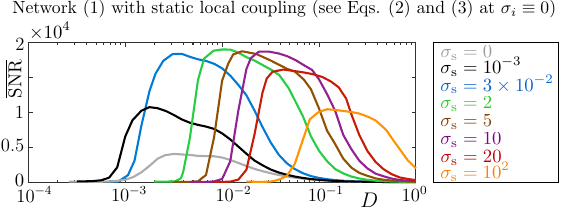}
\caption{Evolution of the curves $\overline{\text{SNR}}(D)$ in network (\ref{eq:general}) with local static coupling  (see Eqs. (\ref{eq:local_individual}) and Eqs. (\ref{eq:local_summary}) at $\sigma_{i}\equiv 0$ and $\sigma_{\text{s}}\geq 0$) when increasing coupling strength $\sigma_{\text{s}}$. The rest of the parameters are the same as in previous figures.}
\label{fig5}
\end{figure}

The evolution of the $\overline{\text{SNR}}(D)$ curves shown in Fig.~\ref{fig3} can be attributed to the transformation of the oscillations $\sigma_i(t)$ with varying adaptivity parameter $\gamma$ and noise intensity $D$. As it can be seen in Fig.~\ref{fig4}~(a), the mean value of $\overline{\sigma}(t)$ increases significantly with increasing noise intensity, whereas the oscillation amplitude remains approximately unchanged. Furthermore, increasing the parameter $\gamma$ at any noise intensity results in higher mean values $<\overline{\sigma}(t)>$ [Fig.~\ref{fig4}~(b),(c)]. Thus, the response of the network to an increase in $\gamma$ can be interpreted as growth of the static component of the coupling strength. This explains why the transformations of the $\overline{\text{SNR}}(D)$ curves in Fig.~\ref{fig3} reproduce the effects associated with increasing the strength of static local coupling (compare Fig.~\ref{fig3} and Fig.~\ref{fig5}). It is important to note that there is no contradiction in the results illustrated in Fig. \ref{fig5} and the materials reported in Ref.~\cite{semenov2022} where the local-coupling-induced enhancement of stochastic resonance is shown to have a saturable character: it achieves maximum at certain coupling strength and then no visible changes occur except of horizontal shift of the $\overline{\text{SNR}}(D)$ curves (see Fig.~1~(g) in Ref.~\cite{semenov2022}). However, the coupling strength range considered in Ref. ~\cite{semenov2022} is much smaller as compared to the current research and does not include the area of extremely strong coupling suppressing the regularity of the noise-induced oscillations.

%%%%%%%%%%%%%%%%%%%%%%%% FIG 4 %%%%%%%%%%%
%%%%%%%%%%%%%%%%%%%%%%%%%%%%%%%%%%%%%%

%%%%%%%%%%%%%%%%%%%%%%%% FIG 5 %%%%%%%%%%%
%%%%%%%%%%%%%%%%%%%%%%%%%%%%%%%%%%%%%%

\subsection{Global coupling}
\label{sec_phenom_global}
\begin{figure}
\centering
\includegraphics[width=1.00\textwidth]{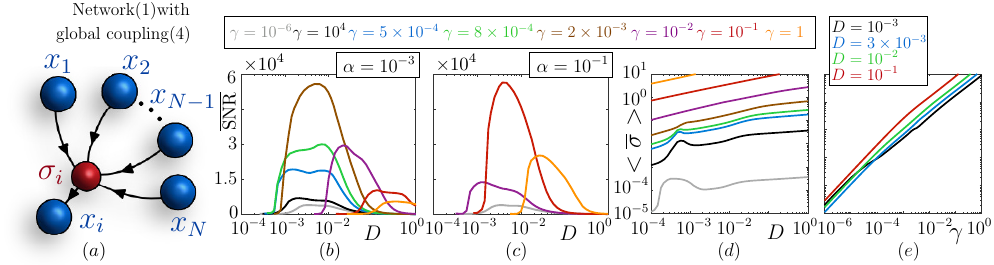}
\caption{Evolution of the stochastic resonance dynamics in network (\ref{eq:general}) with global adaptive coupling governed by Eqs. (\ref{eq:global}):
(a) Schematic illustration of the global coupling structure; (b)-(c) Curves $\overline{\text{SNR}}(D)$ when varying adaptivity parameter $\gamma$ for fixed coupling decay parameter $\alpha=0.001$ (panel (b)) and $\alpha=0.1$ (panel (c)); (d)-(e) Mean value of oscillatory process $\overline{\sigma}(t)$ as a function of noise intensity $D$ (panel (d)) at different fixed $\gamma$ (the color scheme is the same as in panels (b) and (c)) and as a function of $\gamma$ at fixed $D$ (panel (e)). Other network's parameters are the same as in the previous figures.}
\label{fig6}
\end{figure}

Next, the impact of global adaptive coupling is studied. The coupling equations represent model (\ref{eq:local_summary}) extended to the global topology [Fig.~\ref{fig6}~(a)]:
\begin{equation}
\label{eq:global}
\begin{array}{l}
f_i(\vec{x},\vec{\sigma})= \dfrac{(\sigma_{i}+\sigma_{\text{s}})}{N} \sum\limits_{j=1}^{N} \Bigl\{ x_j-x_i \Bigr\}, \quad
\dfrac{d\sigma_{i}}{dt}=\dfrac{ \gamma \Delta_i }{ 1+\Delta_i } - \alpha \sigma_i, \quad
\Delta_i=\left| x_i - \dfrac{1}{N}\sum\limits_{j=1}^{N}  x_j  \right|.
\end{array}
\end{equation}
%%%%%%%%%%%%%%%%%%%%%%%% FIG 6 %%%%%%%%%%%
%%%%%%%%%%%%%%%%%%%%%%%%%%%%%%%%%%%%%%
In such a case, increasing the adaptivity parameter $\gamma$ enhances stochastic resonance. Similarly to the system with static coupling only (compare Fig. 1 in Ref. \cite{semenov2026} and Fig.~\ref{fig6}~(b) in the current paper), the stochastic resonance enhancement has a resonant character: there exists an appropriate adaptivity parameter value $\gamma\approx 0.2$ corresponding to the most pronounced stochastic resonance and maximal peak values of SNR [Fig.~\ref{fig6}~(b)]. The stochastic resonance enhancement is accompanied by shift of the optimal noise level corresponding to the most regular dynamics: $D_{\text{opt}}$ varies from $1.4\times 10^{-3}$ at $\gamma=5\times 10^{-4}$ to $3\times10^{-1}$ at $\gamma=1$ in Fig.~\ref{fig6}~(b).  As illustrated in Fig.~\ref{fig6}~(b), global coupling provides for achieving higher peak SNRs in comparison with the local one, but further drop in the peak SNR values is more pronounced. As the decay parameter $\alpha$ growths, the influence of $\gamma$ becomes weaker, while the qualitative changes do not occur [Fig.~\ref{fig6}~(c)]. Similarly to the local coupling, the clue to understand the global coupling impact is hidden in the evolution of oscillations $\overline{\sigma}(t)$, whose mean values increases monotonically, as $\gamma$ growths [Fig.~\ref{fig6}~(d)-(e)], i.e. the network's dynamics evolution can be interpreted as the behaviour in the presence of increasing static coupling component.

%It is important to note that the particular form of the coupling nonlinearity does not play a significant role as long as the coupling satisfies the condition that a larger phase difference corresponds to a stronger interaction. In particular, qualitatively similar effects are observed when the coupling state equation in model (\ref{eq:global}) is replaced by $\dfrac{d\sigma_{i}}{dt}= \gamma \Delta_i^2 - \alpha \sigma_i$. 

\subsection{Simultaneous impact of dynamical and static coupling}
\label{sec_simultaneous}
\begin{figure}
\centering
\includegraphics[width=0.68\textwidth]{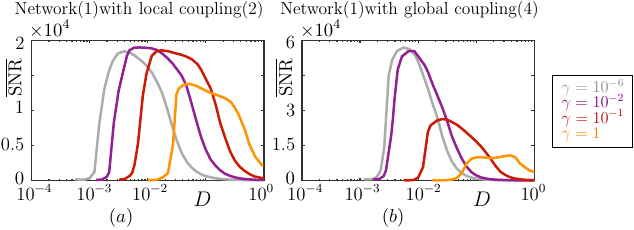}
\caption{Evolution of the stochastic resonance curves $\overline{\text{SNR}}(D)$ in network (\ref{eq:general}) with coupling (\ref{eq:local_individual}) (panel (a)) and (\ref{eq:global}) (panel (b)) when varying adaptivity parameter $\gamma$ at fixed coupling parameters $\alpha=0.001$ and $\sigma_{\text{s}}=0.2$. The rest of the parameters are the same as in the previous figures.}
\label{fig7}
\end{figure}

Suppose that the adaptive coupling strengths $\sigma_i$ evolve at nonzero positive static strength $\sigma_{\text{s}}>0$. Since the $\overline{\text{SNR}}(D)$ curves obtained for networks with either static or adaptive coupling exhibit similar behavior, it can be expected that $\gamma$ and $\sigma_{\text{s}}$ act as two coupling parameters that affect stochastic resonance in a similar manner. To testify this assumption, $\sigma_{\text{s}}$ is fixed at a certain level, $\sigma_{\text{s}}=0.2$, while $\gamma$ is subsequently varied to reproduce effects similar to those observed when further increasing $\sigma_{\text{s}}$ in the absence of adaptive coupling (at $\gamma=0$). In the presence of local coupling (\ref{eq:local_individual}) in network (\ref{eq:general}), growth of $\gamma$ shifts the optimal noise intensity toward higher values [Fig.~\ref{fig7}~(a)] and suppresses stochastic resonance, similarly to the effect of further increasing $\sigma_{\text{s}}$ at $\gamma=0$. The same effects are observed in the case of global coupling (\ref{eq:global}), but the stochastic resonance suppression occurs much faster [Fig.~\ref{fig7}~(b)]. Thus, the hypothesis of similar character of the static and adaptive coupling impact has been confirmed.

%%%%%%%%%%%%%%%%%%%%%%%% FIG 7 %%%%%%%%%%%
%%%%%%%%%%%%%%%%%%%%%%%%%%%%%%%%%%%%%%

\section{Memristive coupling}
\label{sec_memristive}
\begin{figure}
\centering
\includegraphics[width=1.00\textwidth]{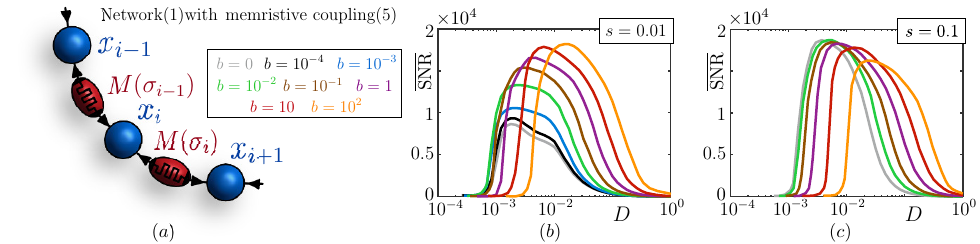}
\caption{Stochastic resonance in network (\ref{eq:general}) with memristive coupling (\ref{eq:memristive}): (a) Schematic illustration of the network; (b)-(c) Evolution of the stochastic resonance curves $\overline{\text{SNR}}(D)$ when increasing the memristivity parameter $b$ at $s=0.01$ (panel (a)) and $s=0.1$ (panel (b)). The memristive forgetting parameter is $\alpha=0.001$. The rest of the network parameters are the same as in the previous figures.}
\label{fig8}
\end{figure}

Consider network (\ref{eq:general}) in the presence of local memristive coupling [Fig.~\ref{fig8}~(a)], whose instantaneous states evolve according to the cubic memristor model \cite{korneev2017}. Following Ref. \cite{korneev2024}, the memristive coupling element model is chosen to be governed by the equations:
\begin{equation}
\label{eq:memristive}
\begin{array}{l}
f_i(\vec{x},\vec{\sigma}) =s \Bigl\{M(\sigma_{i-1})(x_{i-1}-x_i)
+ M(\sigma_i)(x_{i+1} - x_i)\Bigr\},\quad
\dfrac{d\sigma_i}{dt}=x_i-x_{i+1}-\alpha \sigma_i,\quad
M(\sigma_i)=1+b\sigma_i^2, 
\end{array}
\end{equation}
where $s$ denotes the coupling strength, while parameter $b$ characterizes the degree to which the memristive properties are manifested. Parameter $\alpha$ characterizes the memristor forgetting effect associated with the finite correlation between the previous and instantaneous states of the memristor \cite{chang2011,chen2013,zhou2019}. In case $\alpha=0$, memristor-based systems are reported to be exhibiting extreme multistability such that oscillation characteristics continuously depend on initial conditions. In addition to the phenomena mentioned in the last paragraph of the introduction, manifestations of such multistability are bifurcations without parameters \cite{semenov2015,korneev2017,korneev2017-2,korneev2021-2,korneev2023,korneev2023-2} and initial-condition-dependent forced synchronization \cite{korneev2020-2}. As reported in Refs. \cite{semenov2015,korneev2017-2}, physical realization of such effects is problematic due to their structural instability and nonstationarity in the presence of any, even extremely weak, fluctuations and imperfectnesses. For this reason, the limit case $\alpha=0$ is excluded from the consideration. Qualitatively, the evolution of the memristive coupling state is the same as compared to the coupling dynamics discussed above: the strength of interaction increases as the phase difference growths. In case $b=0$ this effect vanishes and the coupling becomes static.

Increasing the memristivity parameter $b$ can give rise to opposite effects depending on the coupling strength $s$. In particular, stochastic resonance is enhanced with increasing $b$ at relatively low values of $s$ [Fig.~\ref{fig8}~(b)]. In contrast, growth of $b$ over the same range of values suppresses stochastic resonance at higher $s$ [Fig.~\ref{fig8}~(c)]. In both cases, the optimal noise intensity corresponding to the peak regularity shifts toward higher value. Thus, the memristive coupling remains to be a factor for controlling stochastic resonance similarly to the phenomenological coupling models discussed above despite of different forms of the coupling state equations. However, there is a principal difference: when increasing $s$ in Eqs. (\ref{eq:memristive}), it impacts both static and dynamic coupling components. Consequently, the stochastic resonance enhancement in Fig.~\ref{fig8}~(b) is achieved under condition of weak static coupling, whereas the suppression in Fig.~\ref{fig8}~(c) occurs in the presence of strong static coupling. In this respect, results for memristively coupled oscillators are consistent with those obtained for the coupling models considered above.

%%%%%%%%%%%%%%%%%%%%%%%% FIG 8 %%%%%%%%%%%
%%%%%%%%%%%%%%%%%%%%%%%%%%%%%%%%%%%%%%

\section{\label{sec:conclusion}Conclusion}
Stochastic resonance is shown here to be efficiently controlled by means of adaptive coupling of different topology and nonlinearity. In the case of phenomenological coupling models characterised by stronger interactions at larger differences between the oscillator states, all of the effects induced by increasing the adaptivity parameters are similar to those reported for the same network in the presence of static coupling only (see Refs. \cite{semenov2022,semenov2026} and Fig.\ref{fig5} of the present paper). In particular, the adaptive-coupling-induced stochastic resonance enhancement has a resonant character: stochastic resonance is first enhanced and then suppressed as  the adaptivity parameter increases, i.e. there exists an optimal parameter value corresponding to the most pronounced stochastic resonance. In the case local coupling, the suppression is observed at high values of the adaptivity parameter values. In contrast, global coupling yields higher peak SNRs when enhancing stochastic resonance, but further suppression occurs more rapidly. For all considered coupling configurations, the dynamic coupling strengths evolve such that their mean value varies significantly, while the amplitude of the coupling oscillations remains relatively small. Thus, the network's structure can be roughly considered to be static. This provides an explanation for the similarity between the collective dynamics of networks with adaptive and static coupling.

In the presence of both adaptive and static coupling components, increasing the static coupling strength and the adaptivity parameter produces similar effects on stochastic resonance. This makes it possible to suppress stochastic resonance by varying the adaptivity parameter when stochastic resonance is already enhanced by a sufficiently strong static coupling. In the case of memristive coupling, where static and adaptive (memristive) components arise naturally, the effects induced by strengthening the memristive properties depend on the coupling strength, which determines the total interaction intensity, including both static and adaptive components. In particular, stochastic resonance can be either suppressed or enhanced at low and high values of the total interaction strength, respectively.

Here, different types of adaptive coupling, characterized by different topologies and nonlinearities, are considered within a unified framework to demonstrate that adaptive coupling can be effectively employed to control stochastic resonance alongside other available approaches. The effect of adaptive coupling on stochastic resonance is undoubtedly determined to a large extent by the intrinsic properties of the particular coupling model. Nevertheless, adaptive coupling provides an effective means of controlling the stochastic resonance response and shifting the noise intensity at which the maximum regularity is achieved.

%Thus, the further studies will be focused on attempts to reveal theoretical reasons for stochastic resonance control to develop applicable requirements to enhancement or suppression of stochastic resonance.

\section*{Declaration of Competing Interest}
The author declares that he has no known competing financial interests or personal relationships that could have appeared to influence the work reported in this paper.

\section*{Data Availability}
The data that support the findings of this study are available from the corresponding author upon reasonable request.

\section*{Acknowledgements}
This work was supported by the Russian Science Foundation (project No. 26-72-10041).

\text{https://rscf.ru/project/26-72-10041/}

%%% Loading bibliography style file
%\bibliographystyle{model1-num-names}
%%\bibliographystyle{cas-model2-names}
%
%% Loading bibliography database
%%\bibliography{bibliography}

\end{document}